\documentclass[letterpaper,twocolumn,10pt]{ieeetran}
\usepackage{tikz}
\usepackage{algorithm}
\usepackage{algpseudocode}
\usepackage{amsmath}
\usepackage{graphicx}
\usepackage{comment}
\usepackage{array,multirow,graphicx}
 \usepackage{float}
 \usepackage{amssymb}
 \usepackage{booktabs}
\usepackage{enumitem}
\usepackage{multicol}
\usepackage{makecell}
\usepackage{filecontents}
\usepackage{caption}  
\usepackage{subcaption}
\usepackage{longtable}
\usepackage{balance}
\algnewcommand\algorithmicforeach{\textbf{for each:}}
\algnewcommand\ForEach{\item[ \algorithmicforeach]}

\begin{document}

\title{Exploring Wireless Channels in Rural Areas: \\A Comprehensive Measurement Study}

 \author{\IEEEauthorblockN{Tianyi Zhang, Guoying Zu, Taimoor Ul Islam, Evan Gossling, Sarath Babu, Daji Qiao, Hongwei Zhang}
 
 \IEEEauthorblockA{Department of Electrical and Computer Engineering, Iowa State University, U.S.A.\\
 \{tianyiz, gyzu, tislam, evang, sarath4, daji, hongwei\}@iastate.edu
 \vspace{-.3in}
 }}
\maketitle
\thispagestyle{empty}

\begin{abstract}
 The study of wireless channel behavior has been an active research topic for many years. However, there exists a noticeable scarcity of studies focusing on wireless channel characteristics in rural areas. With the advancement of smart agriculture practices in rural regions, there has been an increasing demand for affordable, high-capacity, and low-latency wireless networks to support various precision agriculture applications such as plant phenotyping, livestock health monitoring, and agriculture automation. To address this research gap, we conducted a channel measurement study on multiple wireless frequency bands at various crop and livestock farms near Ames, Iowa, based on Iowa State University~(ISU)'s ARA Wireless Living lab - one of the NSF PAWR platforms. We specifically investigate the impact of weather conditions, humidity, temperature, and farm buildings on wireless channel behavior. The resulting measurement dataset, which will soon be made publicly accessible, represents a valuable resource for researchers interested in wireless channel prediction and optimization.
 %Observations from this study provide valuable insights into atmospheric attenuation and shadowing in rural areas, thereby enhancing our understanding of wireless channels in agricultural environments.
\end{abstract}

%\vspace{-0.6cm}

%\keywords{ARA, rural wireless, channel measurement, }


\maketitle
\vspace{-0.5cm}
\section{Introduction} \label{sec:Introduction}
%In recent years,
%With the rapid advancement of technology, 
Numerous cutting-edge technologies are being integrated into agricultural practices, revolutionizing precision and automation in the field. These advancements heavily rely on the use of cameras, sensors, and unmanned vehicles to optimize agricultural processes. As demand for the use of such devices in agriculture continues to grow, it has become imperative to ensure reliable wireless communication over a wireless network that possesses characteristics such as large-scale coverage, affordability, high capacity, and low latency. A reliable wireless connection depends not only on advanced hardware and algorithms but also on a thorough understanding of the wireless channel behaviors. Some studies such as the one by Wang et al. in~\cite{wang2018survey} aims to measure and model the behavior of 5G wireless channels. However, agricultural areas are located in rural regions, where the wireless channel characteristics may significantly differ from urban and suburban areas. 

In recent years, wireless sensor networks have gained significant attention in agriculture, leading to numerous studies focusing on measuring wireless channels in various agricultural scenarios. For instance, Jawad et al. derived empirical path loss models in farm fields using drones~\cite{jawad2019accurate}. In a similar way, measurements were conducted in apple orchards in~\cite{abouzar2016rssi} and~\cite{guo2012propagation}. Raheemah et al. generated empirical path loss models in~\cite{raheemah2016new} for greenhouses. Measurements from cornfields were taken by Pan et al. in~\cite{pan2017modeling}, while Zhu et al. conducted measurements in a pig breeding farm~\cite{zhu2017modeling}. However, these studies primarily focused on Zigbee at 2.4\,GHz and their path loss models focus on the impact of distance on the wireless channels.
%addressed distances.

As discussed in~\cite{luomala2015effects} and~\cite{czerwinski2017path}, and further verified in our previous work~\cite{sander2021measurement}, weather conditions have a significant impact on wireless channels. In our previous study, we measured the wireless channels of  TV white space (TVWS) in a crop farm and found that channel quality %performance 
varies considerably between morning and mid-day. However, due to the hardware limitations, we could not collect accurate weather information to establish the quantitative relationship between channel quality and weather conditions. Now, with the ARA wireless living lab~\cite{zhang2022ara}, we have the ability to collect and analyze wireless channel information with the help of accurate and comprehensive weather data.

ARA~\cite{zhang2022ara} as part of the NSF Platforms for Advanced Wireless Research~(PAWR) program, is an at-scale platform for advanced wireless research deployed across the Iowa State University (ISU) campus, City of Ames, Iowa, USA, surrounding research and producer farms, and rural communities in central Iowa, spanning a rural area with a diameter of over 60\,km. ARA serves as a wireless living lab for smart and connected rural communities, facilitating the research and development of rural-focused wireless technologies that provide affordable, high-capacity connectivity to rural communities and industries such as agriculture.

Leveraging the ARA wireless living lab, we conducted a measurement study between March and June of 2023 to analyze the TVWS and mid-band wireless channels in various crop and livestock farms. 
%Our study involved the utilization of an ARA base station (BS), a few fixed-location ARA user equipment (UE), and a portable ARA UE.
%This extensive measurement study encompassed multi-band channels and considered a range of environmental factors. 
Key contributions of this measurement study are summarized below:
\begin{itemize}
    \item We performed a comprehensive analysis of multi-band wireless channels using wireless channel measurement data collected by ARA base stations~(BSs) and user equipment~(UEs), and weather data collected by the weather station and the disdrometer. The weather dataset includes information such as rain rate, raindrop size, humidity, and temperature. %The availability of this data 
%    The results provide a unique perspective on wireless channel behaviors in agricultural settings.
    \item We gathered and analyzed path loss information from different types of building blockages in various crop and livestock farms, which could be valuable for both radio deployments and algorithm designs in the rural settings.
    \item We will make the dataset publicly available that contains time-stamped wireless channel measurements and weather information,
%    all the raw data of channel and weather information with time stamps, 
including the channel matrix of a MIMO system, which could be beneficial to data-driven wireless communications research. %and algorithm design. 

%collected during the measurement campaign, which would be a good support for any data-driven research or algorithm testing.
%We offer a valuable dataset to the public, comprising all the raw channel and weather data with time stamps during our measurement campaign. This dataset also encompasses the MIMO system's channel matrix. Researchers and algorithm developers will find this dataset beneficial for data-driven research and algorithm testing purposes.
\end{itemize}

The rest of this paper is organized as follows: Section~\ref{sec:SystemOverview} presents an overview of the entire system. %Subsequently, 
Section~\ref{sec:Methodology} discusses the methodology employed in this measurement study. We present and analyze the measurement results in Section~\ref{sec:Measurements&results} and section~\ref{sec:Conclusion} summarizes the key findings.
\vspace{-0.25cm}

\section{System overview} \label{sec:SystemOverview}
%Fig: Map with BS and UE locations

ARA has four outdoor base stations~(BSs) and 13~user equipment~(UEs) at fixed locations in Phase-1. In this paper, our research focuses specifically on the BS on the rooftop of ISU Wilson Residence Hall, that is surrounded by essential application facilities of ARA. To the south, there are ISU dairy farm, sheep teaching farm, and Curtiss crop farm. On the west side, a few City of Ames facilities and several farms are located, while agricultural vehicle operate on the east side of the building.
\vspace{-0.25cm}
\subsection{Base Station}
The rooftop base station at Wilson Hall, located approximately 120\,ft above the ground, is equipped with a comprehensive array of wireless equipment. This includes 1$\times$Skylark TVWS BS, 3$\times$Ericsson mid-band BSs, 3$\times$Ericsson mmWave BSs, 3$\times$NI N320 Software-defined radios~(SDRs), 1$\times$Keysight RF Sensor, 1$\times$weather station, and 1$\times$disdrometer. 
The BS consists of three sectors with azimuths of 60, 180, and 300 degrees, and each sector covers 120~degrees. Fig.~\ref{Antennas for BS} shows the
antenna layout of the northwest sector (an azimuth of 60 degrees).
\iffalse
\begin{figure}[htb]
\centering 
\setlength{\abovecaptionskip}{0.cm}
\includegraphics[width=0.4\textwidth]{Figs/wilsonhallantenna.jpg}
\caption{ISU Wilson Hall with BS on the rooftop.}
     \label{WislsonHall}
\end{figure}
\fi

\begin{figure*}[htbp]
\vspace{-0.3cm}
    \begin{minipage}{0.32\textwidth}
        \centering
        \includegraphics[scale=0.1]{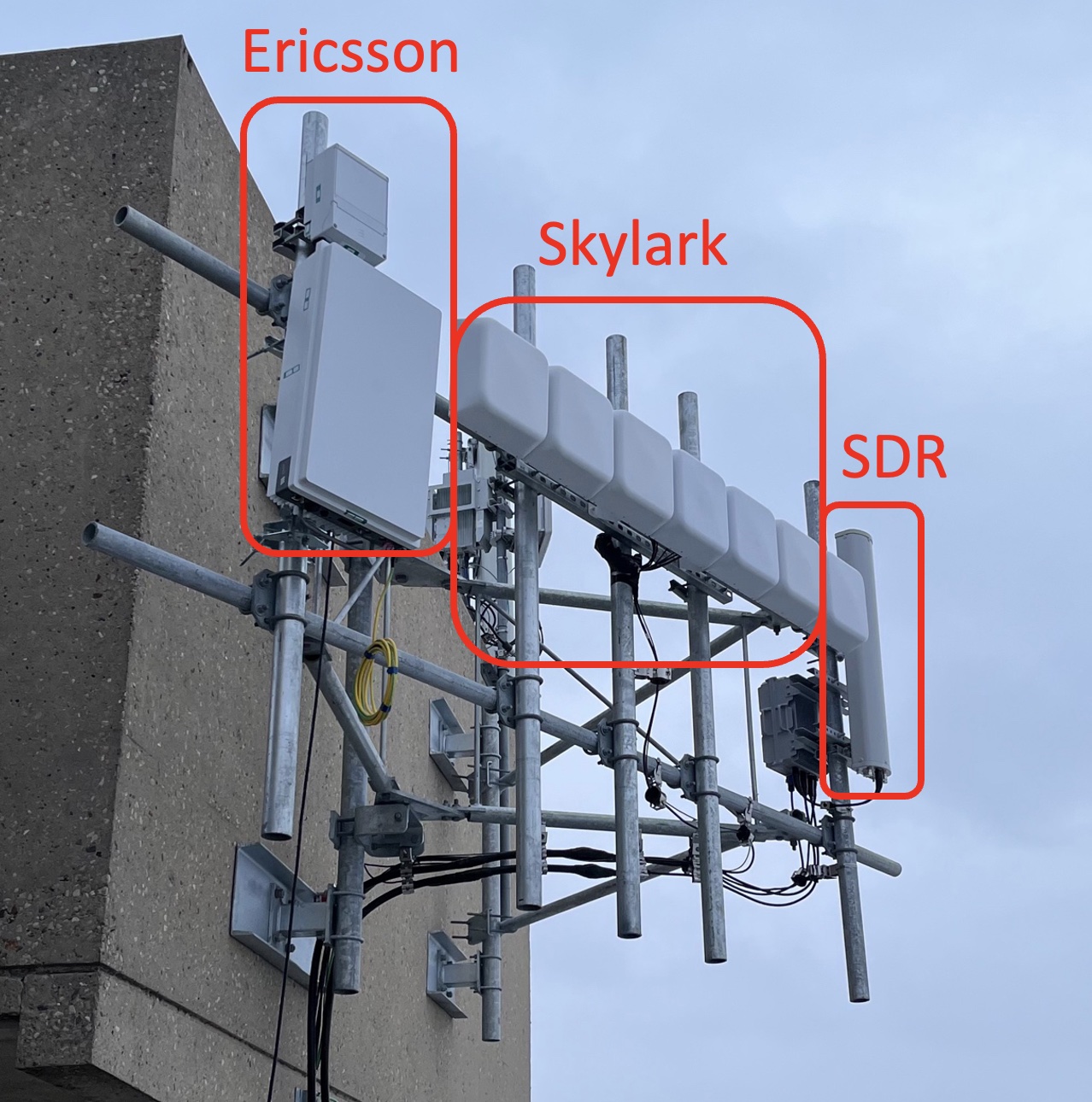}
\caption{Antenna layout of the northwest sector.}
     \label{Antennas for BS}
    \end{minipage}
    \hfill
    \begin{minipage}{0.32\textwidth}
        \centering
        \includegraphics[width=\linewidth]{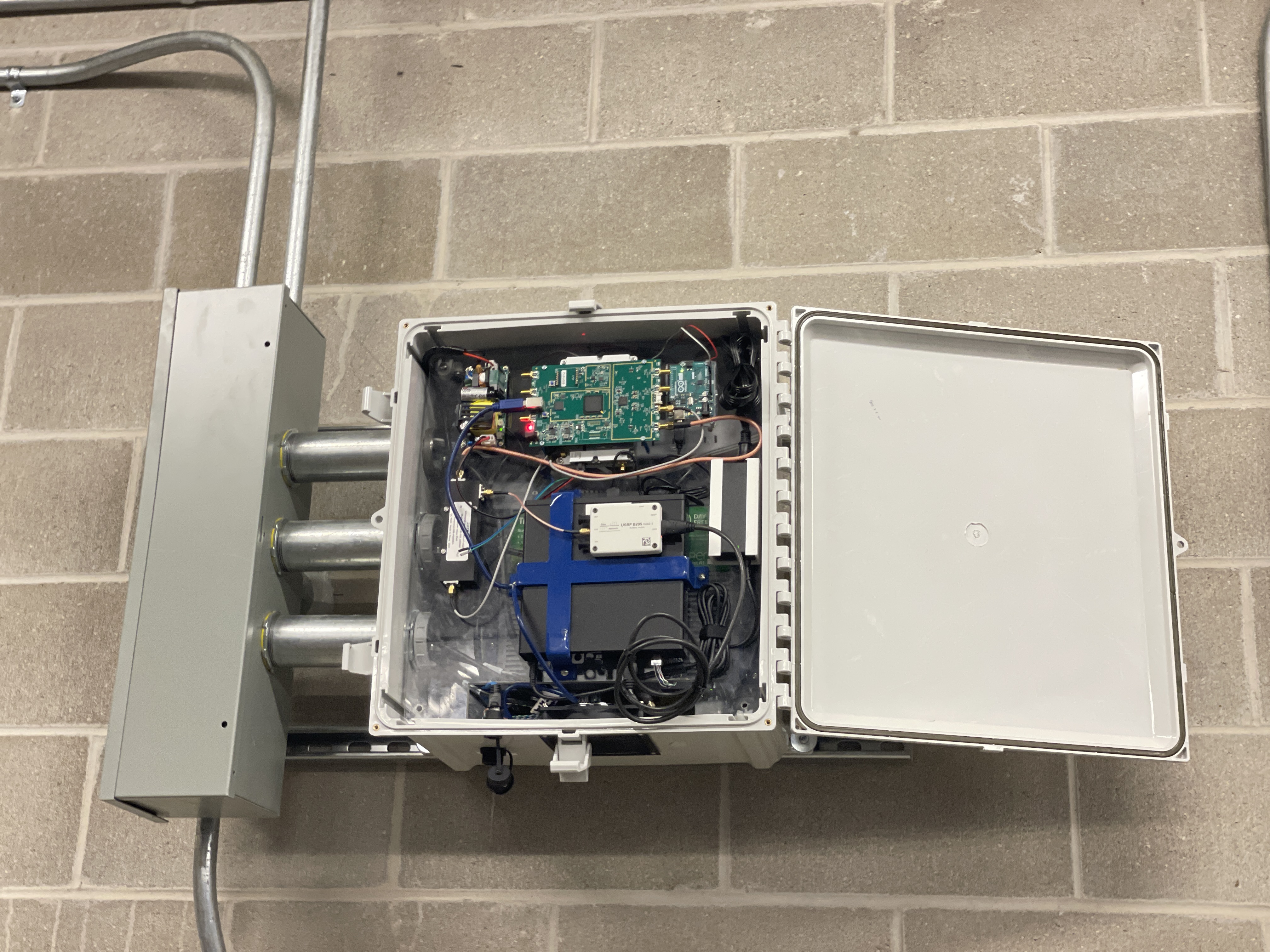}
\caption{An ARA UE box deployed at a fixed location.}
     \label{ueBox}
    \end{minipage}
    \hfill
    \begin{minipage}{0.32\textwidth}
        \centering
        \includegraphics[scale=0.47]{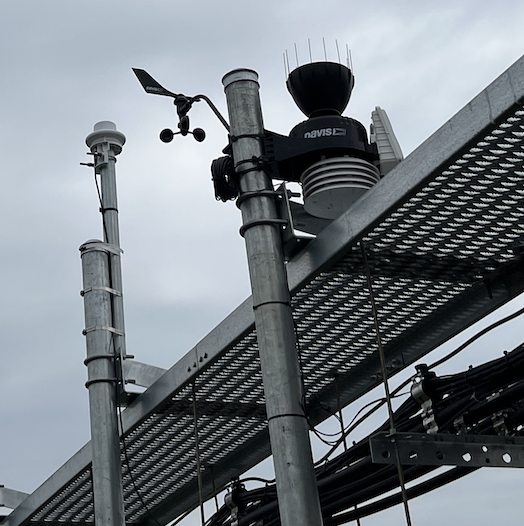}
\caption{Davis weather station (right) and WS100 disdrometer (left).}
     \label{weatherStation}
    \end{minipage}
    \vspace{-0.6cm}
\end{figure*}
\iffalse
\begin{figure}[htb]
\centering 
\setlength{\abovecaptionskip}{0.cm}
\includegraphics[width=0.3\textwidth]{Figs/wilsonhallantenna.JPG}
\caption{Antenna layout of the northwest sector.}
     \label{Antennas for BS}
\end{figure}
\fi

The Skylark BS is a commercial-off-the-shelf (COTS) platform designed to operate in the TVWS bands. It comprises one central unit (CU), one distributed unit (DU), and three radio units (RUs). Each RU is equipped with 14 antennas.
%, facing northwest, northeast, and south individually. 
Skylark supports many-antennas multiple-input multiple-output~(mMIMO) technology and is poised to support Open-RAN in the future. The Skylark deployment as part of the ARA wireless living lab, named \emph{AraMIMO}~\cite{aramimo-wintech2023}, provides a set of APIs, enabling control, configuration and measurement, thereby facilitating comprehensive research on whole-stack mMIMO systems.

In the mid-band frequency range, ARA is equipped with three Ericsson AIR6419 BSs operating in the range of \mbox{3450--3550\,MHz}. These BSs support mMIMO as well as CSI-RS and SRS beam~forming. ARA has also deployed three NI USRP N320 SDRs operating in the mid-band. SDRs are programmable transceivers that offer flexible, reconfigurable, and programmable framework for various wireless technologies, eliminating the need for hardware updates. In ARA, SDRs are integrated with power amplifiers~(PA) and low-noise amplifiers~(LNA) to enhance signal strength in outdoor environments. %the amplifiers limit 
These amplifiers operate between 3400--3800\,MHz. The SDRs can be controlled using USRP Hardware Driver~(UHD)~\cite{UHDEttus98:online} or GNURadio~\cite{GNURadio99:online}, enabling functions such as spectrum sensing, signal generation and analysis, as well as running the full-stack LTE and 5G using open-source software such as srsRAN~\cite{srsRANPr74:online} and OpenAirInterface~\cite{OpenAirI26:online}.
\vspace{-0.2cm}
\subsection{User Equipment}

ARA is equipped with two types of UEs: fixed-location UE and portable UE. Fixed-location UEs are strategically placed in crop and livestock farms to facilitate agricultural and livestock sciences research, while portable UEs are designed to be mounted on various vehicles used for agriculture, school and public transport, and fire and safety services. In this measurement study, portable UEs are used to assess how wireless links are affected by their surrounding environments, e.g., blockage characteristics of various farm buildings.

As depicted in Fig.~\ref{ueBox}, % and \ref{ueArch},
each UE is housed within a box that consists of a Skylark customer premises equipment~(CPE) operating in the TVWS band, a Quectel UE to communicate with Ericsson BS in both mid-band and mmWave band, and an NI B210 SDR operating in the mid-band.
%, each catering to different frequency bands. 
In addition, each UE box is equipped with a Cradlepoint IBR600C router, enabling management and provisioning of the UE devices through the ARA portal~\cite{loginOpenSt43:online}.
%The Skylark CPE is equipped with two antennas, which work in conjunction with the Skylark BS. For communication with the Ericsson BS, the 5G module of RG530 from Quectel is deployed, operating within the Sub-7GHz frequency band. Furthermore, each box includes one NI USRP B210 and one NI USRP B205-mini SDR.

%missing citation: srs skylark oai uhd gnuradio
\iffalse
\begin{figure}[htb]
\centering 
\setlength{\abovecaptionskip}{0.cm}
\includegraphics[width=0.3\textwidth]{Figs/UEBox.JPG}
\caption{An ARA UE box deployed at a fixed location.}
     \label{ueBox}
\end{figure}
\fi
%\begin{figure}[!htb]
%\centering 
%\setlength{\abovecaptionskip}{0.cm}
%\includegraphics[width=0.4\textwidth]
%{Figs/UE_Architecture.png}
%\caption{ARA UE box architecture}
%     \label{ueArch}
%\end{figure}
\vspace{-0.25cm}
\subsection{Weather Station and Disdrometer}

ARA features weather stations and high-precision disdrometers at BS sites~(Fig.~\ref{weatherStation}) to collect weather information. These devices enable continuous collection of weather data such as temperature, humidity, rain rate, and raindrop size. By correlating weather data with channel measurement results, 
%analyzing the data obtained from these devices, 
we could gain a comprehensive understanding of how the environmental variables impact the wireless channel condition. This knowledge serves as a crucial foundation for uninterrupted ultra-reliable low-latency communication (URLLC) under all-day, all-weather conditions.
\vspace{-0.2cm}
\iffalse
\begin{figure}[!htb]
\centering 
\setlength{\abovecaptionskip}{0.cm}
\includegraphics[width=0.3\textwidth]{Figs/weatherStation.png}
\caption{Davis weather station (right) and OTT Parsivel$^2$ disdrometer (left) on the Wilson Hall rooftop.}
     \label{weatherStation}
\end{figure}
\fi

\section{Methodology} \label{sec:Methodology}
The measurement study took place between March and June of 2023 and is divided into two parts: fixed-location UE measurements and portable UE measurements, each serving different purposes.
\vspace{-0.3cm}
\subsection{Fixed-location UE measurements} \label{subsec:FixedUE}
The primary goal of the fixed-location UE measurements was to collect data and study %related to 
the impact of weather conditions. 
%Fixed UEs were chosen for data collection due to their remote controllability. 
Automated scripts were developed and run on both BS and UE host computers for data collection. These scripts allow for the customization of measurement parameters such as starting time, duration, center frequency, and bandwidth. 
%For the purpose of this paper, 
In this paper, we focus our study on 
%the common data that could be collected from both 
Skylark~(TVWS band) and Ericsson/Quectel~(mid-band).
%hardware. 
At the UE side, the scripts were designed to collect only the received signal strength, while on the BS side, the scripts collected throughput, latency, and signal-to-noise ratio~(SNR). Furthermore, the scripts at the BS side initiate the scripts at the UE side to ensure synchronized data collection at both ends.

\begin{figure}[htbp]
\centering 
\setlength{\abovecaptionskip}{0.cm}
\includegraphics[width=0.3\textwidth]{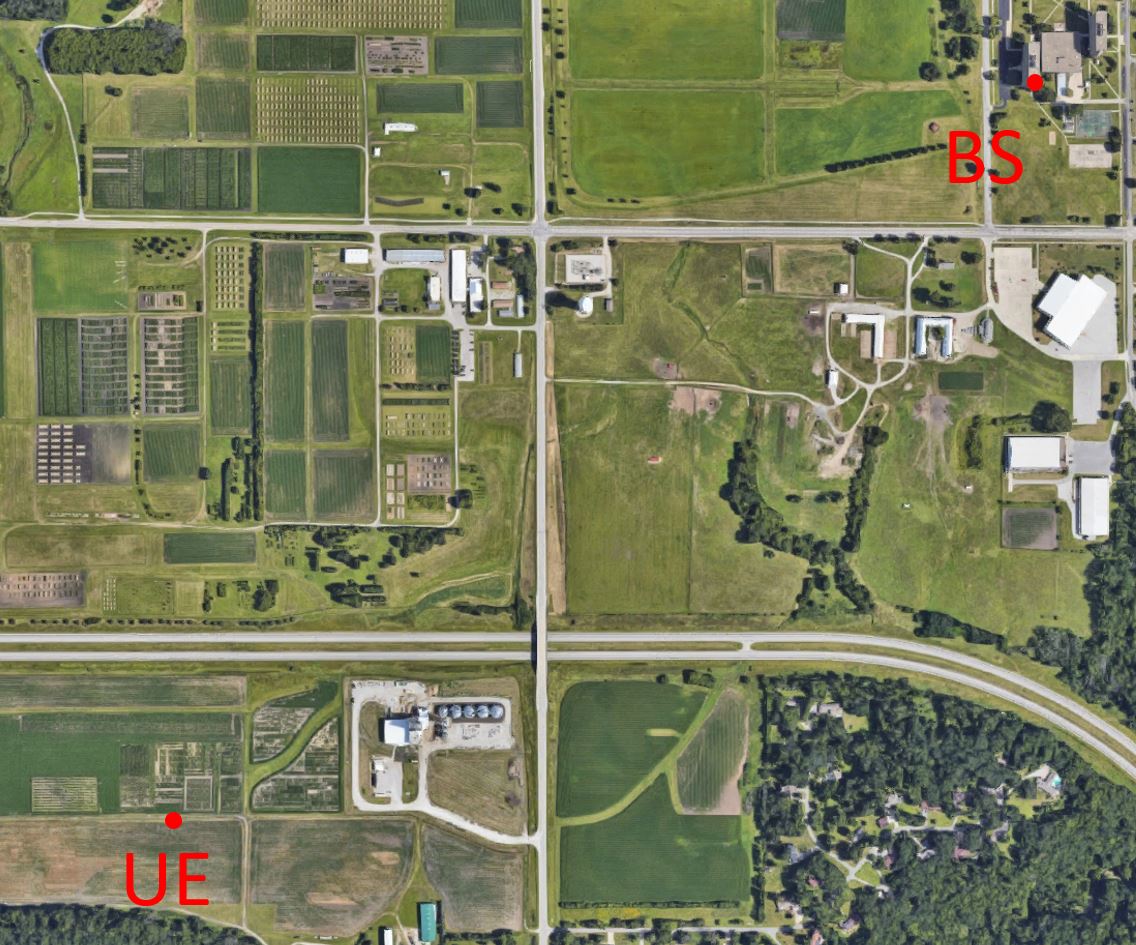}
\caption{A line-of-sight path of 0.94~miles between Wilson Hall BS and a fixed-location UE at the Curtiss Farm field.}
     \label{fixedNode}
     \vspace{-0.2cm}
\end{figure}

In the experiments, we mainly used a fixed-location UE deployed at the Curtiss Farm field and the BS on the rooftop of Wilson Hall. The geographical locations of these nodes are shown in Fig.~\ref{fixedNode}, with a line-of-sight~(LOS) path of 0.94~miles between them,
%A line-of-sight (LOS) path existed between the BS and UE, 
enabling a reliable connection throughout the entire duration of the experiments. This allows for long-term automated measurements without interruption.
%without concern for data loss due to disconnections. 
Data from the TVWS bands is collected at 2~seconds intervals, while for the mid-band, the interval is set to 8~seconds.

%\vspace{-0.2cm}

Previous studies have demonstrated that wireless communication, particularly in higher frequency bands, are susceptible to precipitation~\cite{czerwinski2017path}.
%negatively affected by raindrops and snowflakes 
In this work, we collect various wireless channel parameters %on various response variables, including
such as path loss, SNR, throughput, and latency, under different weather conditions and across different frequency bands. Measurements were taken for several hours to account for varying levels of precipitation. Weather information reported by the weather station and disdrometer was also incorporated to facilitate the study of the impact of different levels of precipitation.

Furthermore, to study the impact of humidity 
%considering the influence of humidity 
on wireless link quality, as acknowledged in the previous studies~\cite{sander2021measurement, luo2011understanding}, we also collect data during various time intervals on sunny days. We recorded humidity data using the weather station simultaneously. %This approach allows us to investigate the impact of humidity on the observed outcomes.
\vspace{-0.2cm}
\subsection{Portable UE measurements}\label{portable}

%To address the limitations of fixed UE, we utilized portable UEs in our study. By employing portable UEs, we were able to change the locations of the UEs and compare performance differences across various positions. 

We used the portable UEs to compare the wireless link performance at different locations. Even though there exist numerous mature channel models describing the change in channel behavior with distance, this paper specifically focuses on studying the blockages caused by buildings in farms, which have not been extensively investigated so far. Farm environments consist of different types of buildings with unique structures designed for specific purposes, such as crop storage, agricultural machinery storage, and sheep/cattle breeding. Such specialized structures cannot be easily modeled using existing models. Therefore, the objective of this section is to fill the gap in knowledge regarding blockages caused by farm buildings.

%\vspace{-0.2cm}
\begin{figure}[htbp]
\centering 
\setlength{\abovecaptionskip}{0.cm}
\includegraphics[width=0.25\textwidth]{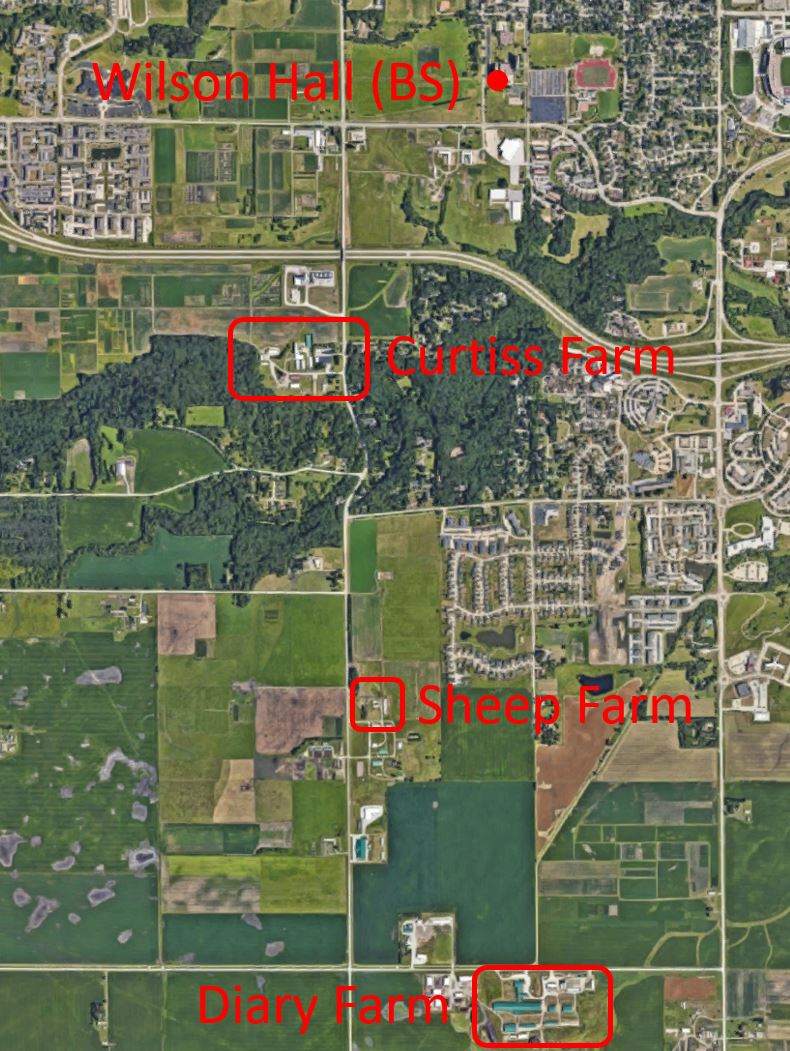}
\caption{Three farms for portable UE measurements.}
     \label{Farms}
     \vspace{-0.25cm}
\end{figure}

\begin{figure}[htbp]
\centering 
\setlength{\abovecaptionskip}{0.cm}
\includegraphics[width=0.25\textwidth]{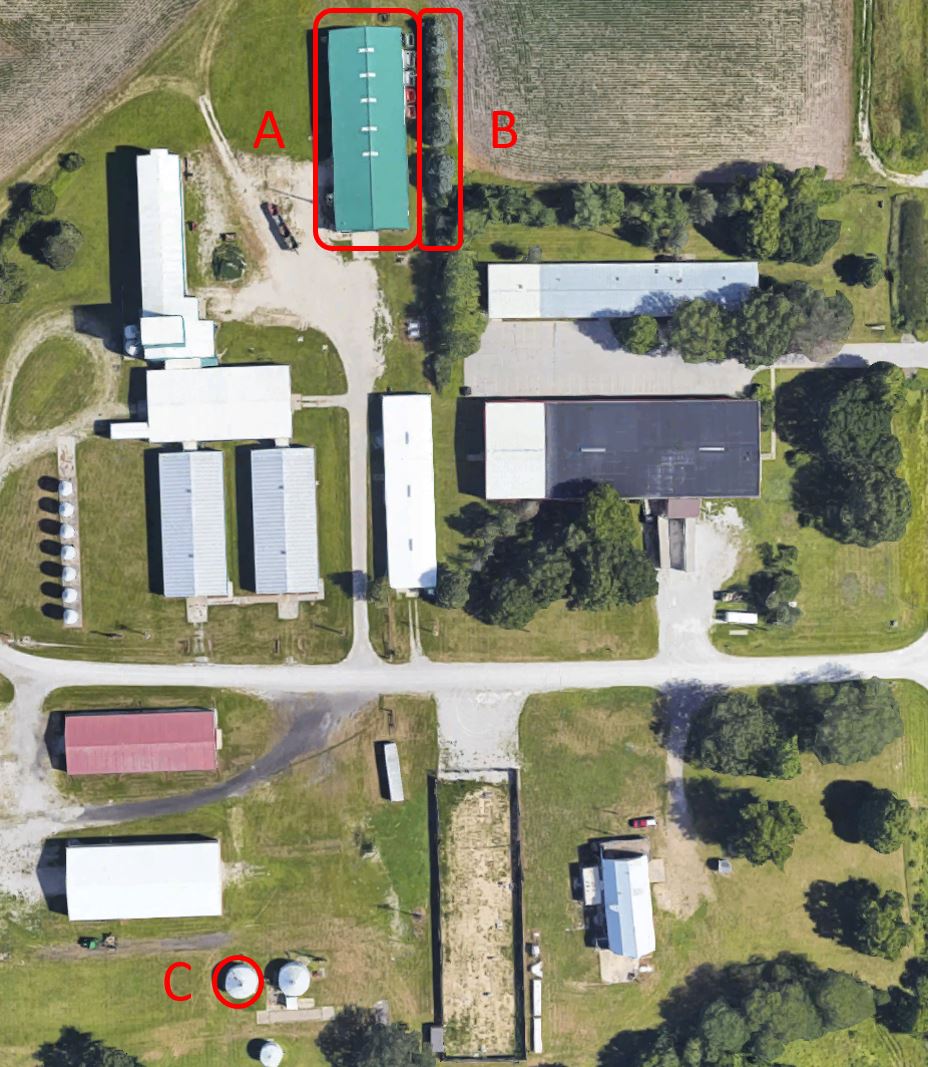}
\caption{Three groups of measurements at Curtiss Farm.}
     \label{Curtiss}
     \vspace{-0.6cm}
\end{figure}

The portable UE measurements were carried out at three farm sites: Curtiss Farm, Sheep Farm, and Dairy Farm, as shown in Fig.~\ref{Farms}.
Curtiss Farm is a crop farm cultivating corn and soybeans. Our measurements at Curtiss Farm were divided into three groups, as shown in Fig.~\ref{Curtiss}. Group~A involves measuring a barn used for agricultural machinery storage, with a large gate facing north. We conducted measurements at five locations, with an additional reference location, under two conditions: (1)~with the gate opened and (2)~with the gate closed. The five locations consist of one situated to the north of the barn, one positioned to the south of the barn, and three locations inside the barn itself, namely inside-north, inside-middle, and inside-south. The reference location is in the nearby field around the same distance toward the Wilson Hall BS, however, with a clear LOS path. This is to 
%was measured at a nearby LOS location, maintaining the same distance, to
facilitate the comparison of the building's impact on wireless links. Groups~B and Group~C focused on measuring the blockage caused by trees and a metal crop storage barn, respectively. Measurements were taken both to the north and the south of the trees/barn.

Similar to Group A measurements at Curtiss Farm, we placed a portable UE to the north and south of the Sheep Farm, as well as three locations inside the building.
The fixed-location UE deployed on the rooftop near the north end of the building is used as the reference node.
%on the rooftop of the Sheep Farm building, where we placed a fixed UE.

\begin{figure}[htbp]
\centering 
\setlength{\abovecaptionskip}{0.cm}
\includegraphics[width=0.4\textwidth]{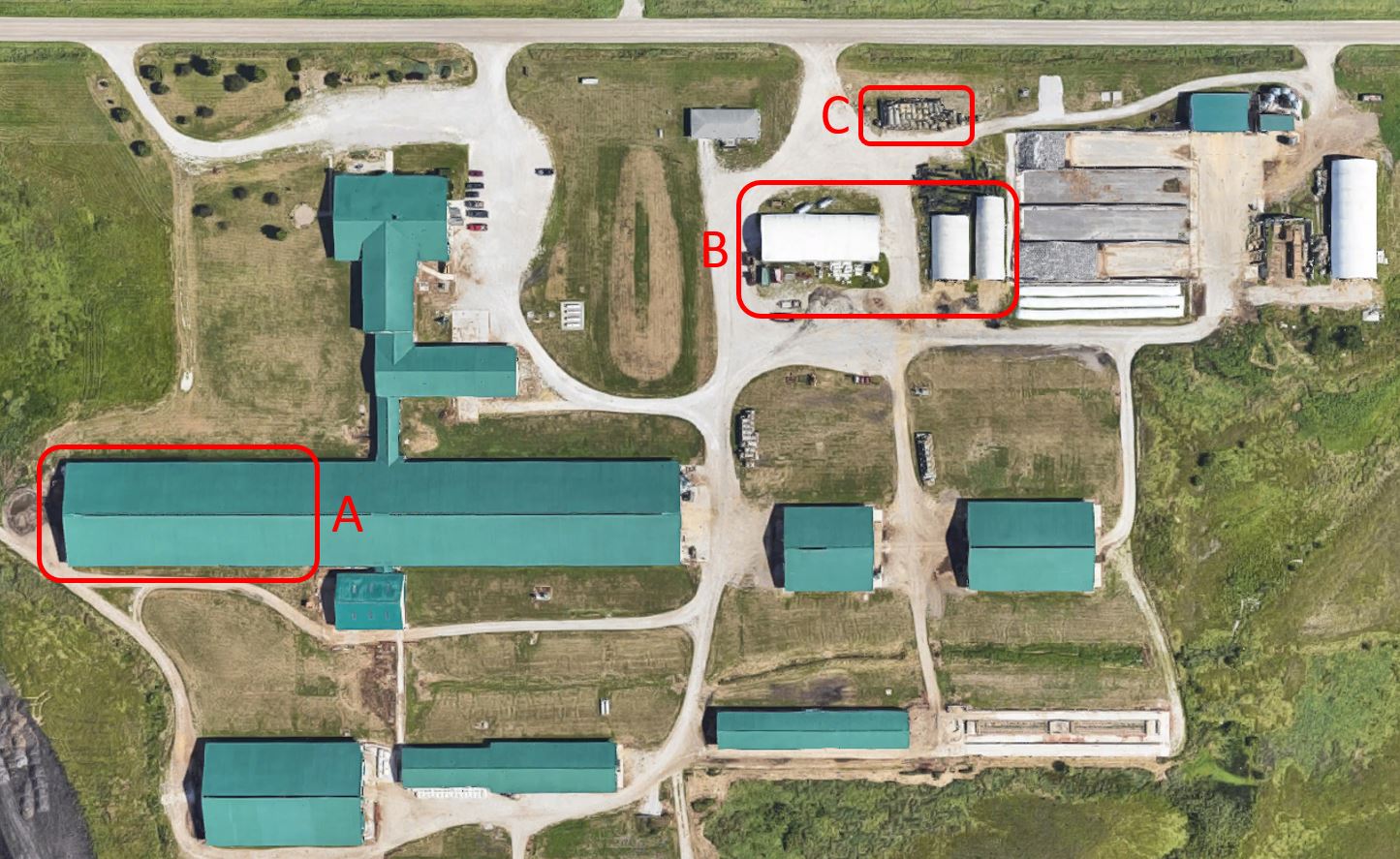}
\caption{Three groups of measurements at Diary Farm.}
     \label{diary}
     \vspace{-0.6cm}
\end{figure}

The measurements at Dairy Farm were also divided into three groups, as depicted in Fig.~\ref{diary}. Group~A involves measurements at the lactation barn. Measurements were taken to the north, to the south, and inside the barn. Group~B includes two hoop houses, one facing the west and another facing the south. We performed the measurements on both hoop houses due to their different orientations. Group~C is dedicated to the measurement of the blockage resulting from a large hay pile. Although hay piles are not as tall as buildings, we include this measurement to account for its potential blockage effects, as hay piles are a common source of blockage in farms.

\vspace{-0.3cm}

\section{Measurement results} 
\label{sec:Measurements&results} 
In this section, we present a subset of our analysis highlighting the practical value of weather information and the distinctive data regarding the impact of farm buildings. Additional data will be made accessible to the public via the ARA data warehouse.
\vspace{-0.4cm}
\subsection{Impact of Rain} \label{subsec:Rain}
\vspace{-0.1cm}
As mentioned in Section~\ref{subsec:FixedUE}, we use a fixed-location UE at Curtiss Farm to evaluate the impact of different weather conditions, including rain. Among the observed weather events, we focus on analyzing the effects of a single rain event in this paper to isolate the influence of other uncontrollable variables. 
%*********************
To gain a better understanding of the impact of rain, we present Figs.~\ref{RainRateE} and \ref{RainRateS} %These figures 
which organize the data by dividing the rain rate into five distinct levels. Higher rain rates result in increased path loss. However, even in the presence of the highest rain rate, the received signal strength only experiences a drop of 1.49~dB in the mid-band and 1.09~dB in the TVWS band when compared to no rain.

\begin{figure*}[htbp]
    %\vspace{-0.3cm}
    \begin{minipage}{0.32\textwidth}
        \centering
        \includegraphics[width=\linewidth]{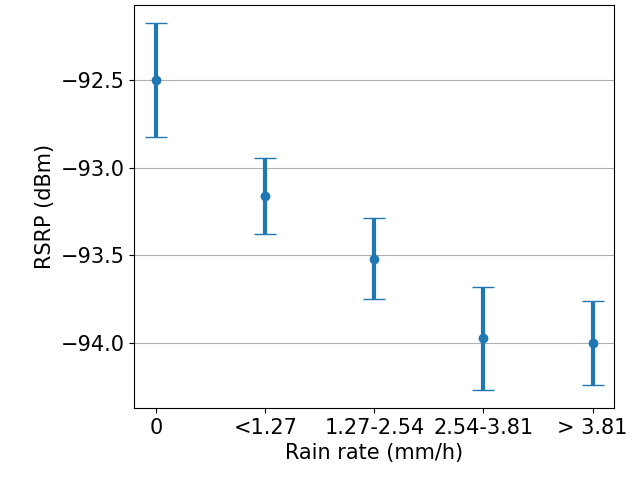}
     \caption{RSRP vs. rain rate in the mid-band with 95\% CI.}
     \label{RainRateE}
    \end{minipage}
    \hfill
    \begin{minipage}{0.32\textwidth}
        \centering
        \includegraphics[width=\linewidth]{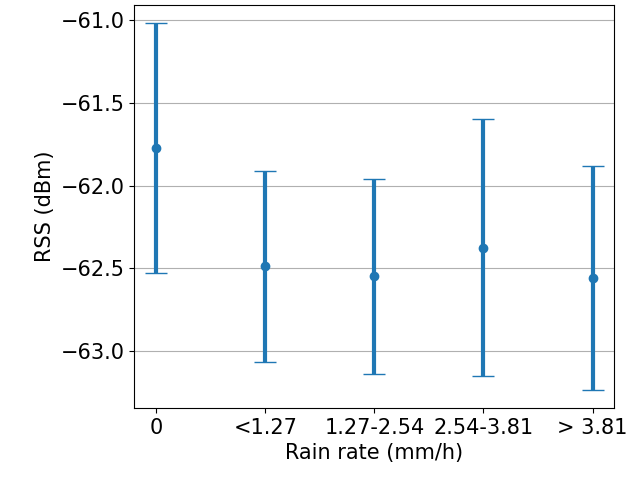}
     \caption{RSS vs. rain rate in the TVWS band with 95\% CI.}
     \label{RainRateS}
    \end{minipage}
    \hfill
    \begin{minipage}{0.32\textwidth}
        \centering
        \includegraphics[width=\linewidth]{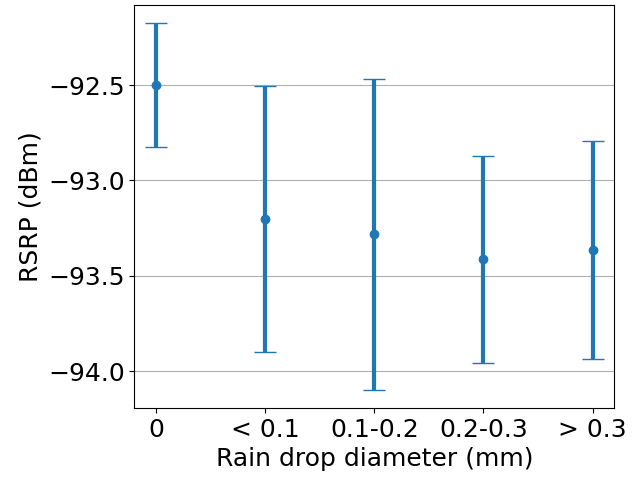}
        \caption{RSRP vs. raindrop diameter in the mid-band with 95\% CI.}
     \label{RaindropE}
    \end{minipage}
    \vspace{-0.3cm}
\end{figure*}

\iffalse%**************************
\begin{figure}[htb]
    \vspace*{-0.1in}
\centering 
\setlength{\abovecaptionskip}{0.cm}
\includegraphics[width=0.4\textwidth]{Figs/rainrateE.png}
\caption{RSRP vs. rain rate in the mid-band with 95\% CI.}
     \label{RainRateE}
    \vspace*{-0.1in}
\end{figure}

\begin{figure}[htb]
\centering 
\setlength{\abovecaptionskip}{0.cm}
\includegraphics[width=0.4\textwidth]{Figs/rainrateS.png}
\caption{RSS vs. rain rate in the TVWS band with 95\% CI.}
     \label{RainRateS}
    %\vspace*{-0.1in}
\end{figure}
\fi%**************************

The ITU-R (International Telecommunication Union - Recommendations) provides a rain attenuation model, P.838-3, in \cite{rec2005itu}, which predicts the attenuation caused by rain based on the rain rate within the frequency range of 1--1,000\,GHz. The specific attenuation, denoted as $\gamma_R$, is determined using a power-law relationship with the rain rate $R$:
\vspace{-0.2cm}
\begin{equation}
    \gamma_R = k R^\alpha,
    \vspace{-0.2cm}
\end{equation}
where coefficients $k$ and $\alpha$ are functions of frequency and can be calculated using Eqns.~(2) and~(3)in~\cite{rec2005itu}. According to this model, the worst-case attenuation caused by rain is estimated to be 0.00226\,dB, which is significantly lower than what we have observed from Figs.~\ref{RainRateE} and \ref{RainRateS}.
One plausible explanation for this substantial attenuation discrepancy is the influence of surface water on the antenna. To gain a deeper understanding of the underlying causes behind this inconsistency, further investigation is warranted. Moreover, there exists potential for the development of a new model designed to account for these variations.

In addition to the rain rate, we also examine the impact of the raindrop size. %However, 
Fig.~\ref{RaindropE} shows limited differentiation in both mean and confidence interval for different raindrop diameters, compared to the previous figures. This lack of distinction could primarily be attributed to the mixing of raindrops with varying diameters in the data. For instance, based on the raw data obtained from the disdrometer, small raindrops (with diameter $\leq$ 1\,mm) are present across all levels of rain rate. Hence, it is not recommended that the raindrop diameter be used as a reliable indicator of the intensity of rain attenuation.
%for rain attenuation.
%not recommended, given the overlapping distribution of raindrops with different diameters.
\iffalse
\begin{figure}[htb]
\centering 
\setlength{\abovecaptionskip}{0.cm}
\includegraphics[width=0.4\textwidth]{Figs/raindropE.png}
\caption{RSRP vs. raindrop diameter in the mid-band with 95\% CI.}
     \label{RaindropE}
\end{figure}
\fi
%Not obvious in latency
%\subsection{Impact of Snow}
%Different from rain, the mid-band data for snow are collected by SDR as Ericsson devices were not ready in the winter. TBD
\vspace{-0.3cm}
\subsection{Impact of Humidity}\label{subsec:humidity}
\begin{figure*}[htbp]
    \begin{minipage}{0.24\textwidth}
        \centering
        \includegraphics[width=\linewidth]{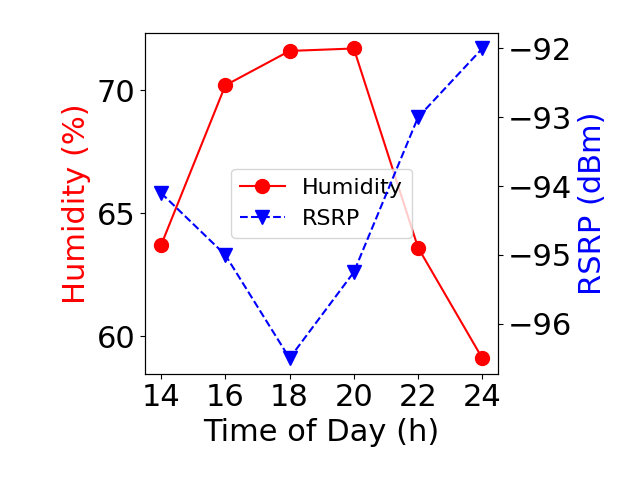}
    \caption{Humidity and mid-band RSRP over time.}
    \label{fig:humidityE}
    \end{minipage}
    \hfill
    \begin{minipage}{0.24\textwidth}
        \centering
        \includegraphics[width=\linewidth]{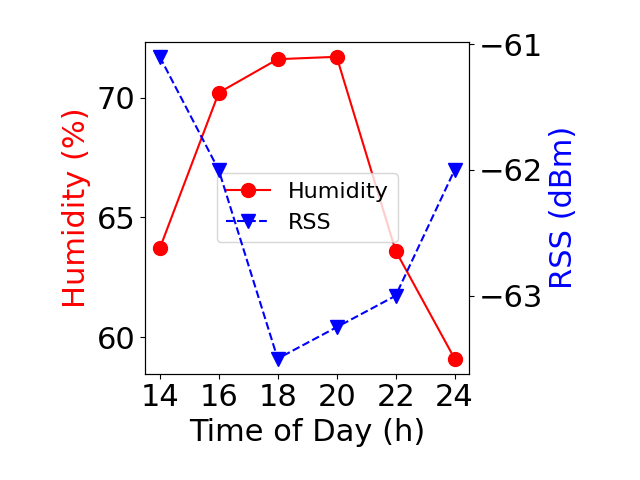}
    \caption{Humidity and TVWS-band RSS over time.}
    \label{fig:humidityS}
    \end{minipage}
    \hfill
    \begin{minipage}{0.24\textwidth}
        \centering
        \includegraphics[width=\linewidth]{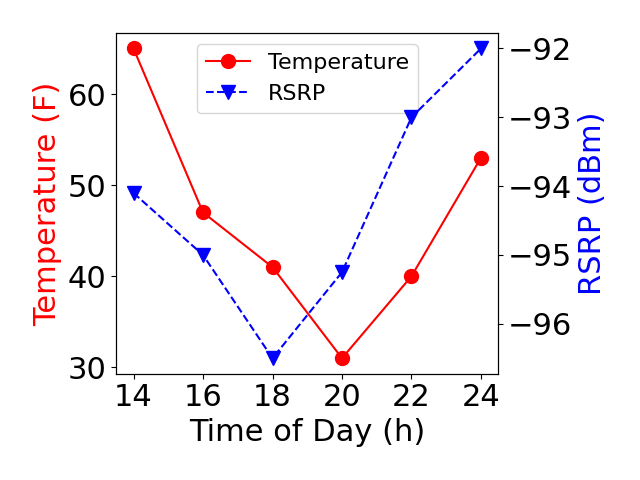}
    \caption{Temperature and mid-band RSRP over time.}
    \label{fig:humidityET}
    \end{minipage}
    \hfill
    \begin{minipage}{0.24\textwidth}
        \centering
        \includegraphics[width=\linewidth]{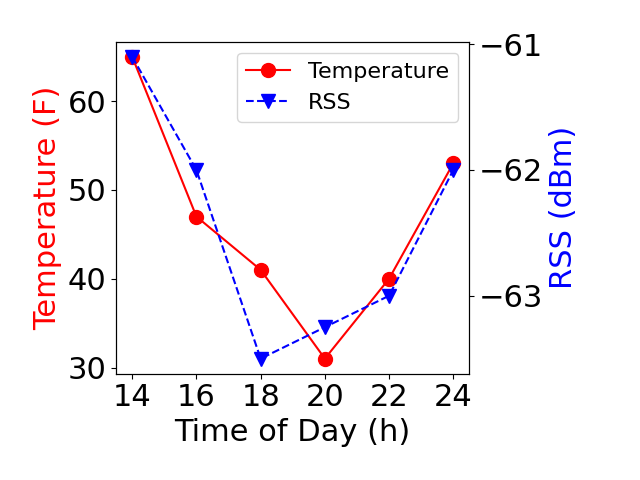}
    \caption{Temperature and TVWS-band RSS over time.}
    \label{fig:humidityST}
    \end{minipage}
    \vspace{-0.3cm}
\end{figure*}

In our previous study in~\cite{sander2021measurement}, we observed a significant impact of time on path loss, with notable variations between morning and mid-day. While we hypothesized that humidity may be a contributing factor, there does not exist sufficient data to validate this hypothesis. In this study, we collected wireless channel information alongside humidity measurements to investigate further. The results, depicted in Figs.~\ref{fig:humidityE} and~\ref{fig:humidityS}, illustrate the influence of humidity on the TVWS and mid-band channels throughout a clear day. Evidently, humidity exhibits an inverse correlation with received signal strength. 
\iffalse
To gain a deeper understanding of the influence of humidity, we have employed the correlation coefficient given by:
\begin{equation}
     r = \frac{{}\sum_{i=1}^{n} (x_i - \overline{x})(y_i - \overline{y})}{\sqrt{\sum_{i=1}^{n} (x_i - \overline{x})^2  \sum_{i=1}^{n}(y_i - \overline{y})^2}},
     \label{eqn:correlation_coefficient}
\end{equation} 
where $x_i$ and $y_i$ represent the received signal strengths in the TVWS and mid-bands, respectively, arranged in chronological order, $n$ denotes the total number of data points for each frequency band, while $\overline{x}$ and $\overline{y}$ indicate the average received signal strengths for the respective bands. The correlation coefficient provides insights into the statistical relationship between the two variables, ranging from -1 to 1. A coefficient of 1 signifies a perfect positive correlation or a direct relationship, while -1 describes a perfect negative or inverse correlation. A correlation coefficient of 0 indicates no linear relationship. 
\fi

To gain a more profound insight into the impact of humidity, we conducted a thorough analysis, calculating the correlation coefficient. This coefficient serves as a valuable tool for assessing the statistical relationship between these two variables, yielding values that span from -1 to 1. A coefficient of 1 signifies a perfect positive correlation, indicating a direct and proportional relationship between the variables, while -1~denotes a perfect negative or inverse correlation. A correlation coefficient of 0, on the other hand, implies the absence of any linear relationship.

In the measurement results of the rain, the correlation coefficient shows a strong negative correlation (-0.94) between the RSRP and humidity in the mid-band, while the \mbox{correlation~(-0.55)} is relatively weaker in TVWS. These observations align with the general expectations, as higher frequency bands tend to experience more signal absorption by water vapor when the frequency is less than 20\,GHz.

\vspace{-0.3cm}
\subsection{Impact of Temperature}\label{subsec:Temperature}
Studies, such as~\cite{luomala2015effects, czerwinski2017path}, believe that temperature is equally important as humidity in affecting wireless channels. In our investigation, we also gathered temperature data to assess its impact. To illustrate our findings, we present the results in Figs.~\ref{fig:humidityET} and~\ref{fig:humidityST}. Both figures demonstrate a positive correlation between temperature and received signal strength. Through correlation coefficient calculations, we found that the TVWS band exhibits a stronger correlation (0.91) with temperature. In contrast, the correlation between temperature and the mid-band is relatively weaker at 0.38. 
We have not yet found a reasonable explanation
%did not uncover any explanation 
for this discrepancy. Considering the insights gained from the humidity analysis in Section~\ref{subsec:humidity}, further investigation is needed to determine whether temperature has a more pronounced impact than humidity in lower frequency bands, and the opposite trend %occurs
in higher frequency bands.
\iffalse
\begin{figure}[htb]
    \centering 
    \setlength{\abovecaptionskip}{0.cm}
    \includegraphics[width=0.4\textwidth]{Figs/humidityET.png}
    \caption{Temperature and mid-band RSRP over time.}
    \label{fig:humidityET}
    \vspace*{-0.05in}
\end{figure}

\begin{figure}[htb]
    \centering 
    \setlength{\abovecaptionskip}{0.cm}
    \includegraphics[width=0.4\textwidth]{Figs/humidityST.png}
    \caption{Temperature and TVWS-band RSS over time.}
    \label{fig:humidityST}
\end{figure}
\fi
\vspace{-0.3cm}
\subsection{Impact of Farm Buidlings}
As discussed in Section \ref{portable}, all measurement locations are to the south of the Wilson Hall BS.
%the BS is situated in the north, while all the buildings are located in the south. Consequently, 
Hence, the signal measurements taken on the north side of the farm buildings were not blocked by any obstacles. In contrast, the signals on the south side of the buildings had to pass through the buildings, resulting in a disparity between the measured signal strengths. Such a discrepancy can be considered as an additional impact caused by the buildings. Moreover, measurements were taken inside the buildings to analyze the path loss due to the walls.

\begin{figure*}
    \vspace{-0.3cm}
    \centering
    \hspace{-2em}%
    \begin{minipage}[htbp]{0.24\textwidth}
    %\begin{subfigure}[!htbp]{0.24\textwidth}
        \centering
        \includegraphics[scale=0.024]{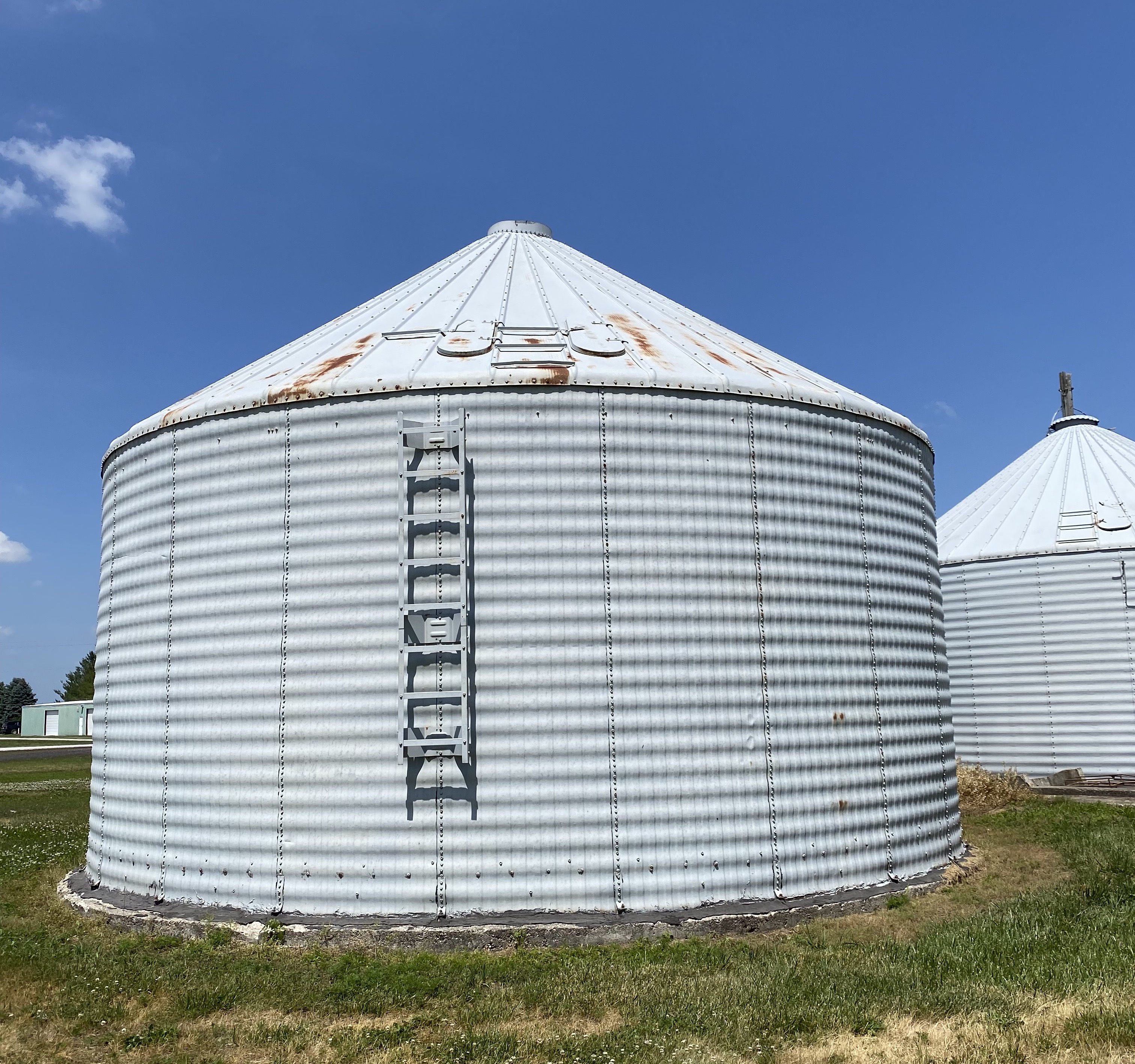}
        \caption{Metal crop barn.}
        \label{fig:barn}
    \end{minipage}
    %\hfill
    %\begin{minipage}{0.24\textwidth}
    %\end{subfigure}
    \hspace{-1em}%
    \begin{minipage}[htbp]{0.24\textwidth}
        \centering
        \includegraphics[scale=0.023]{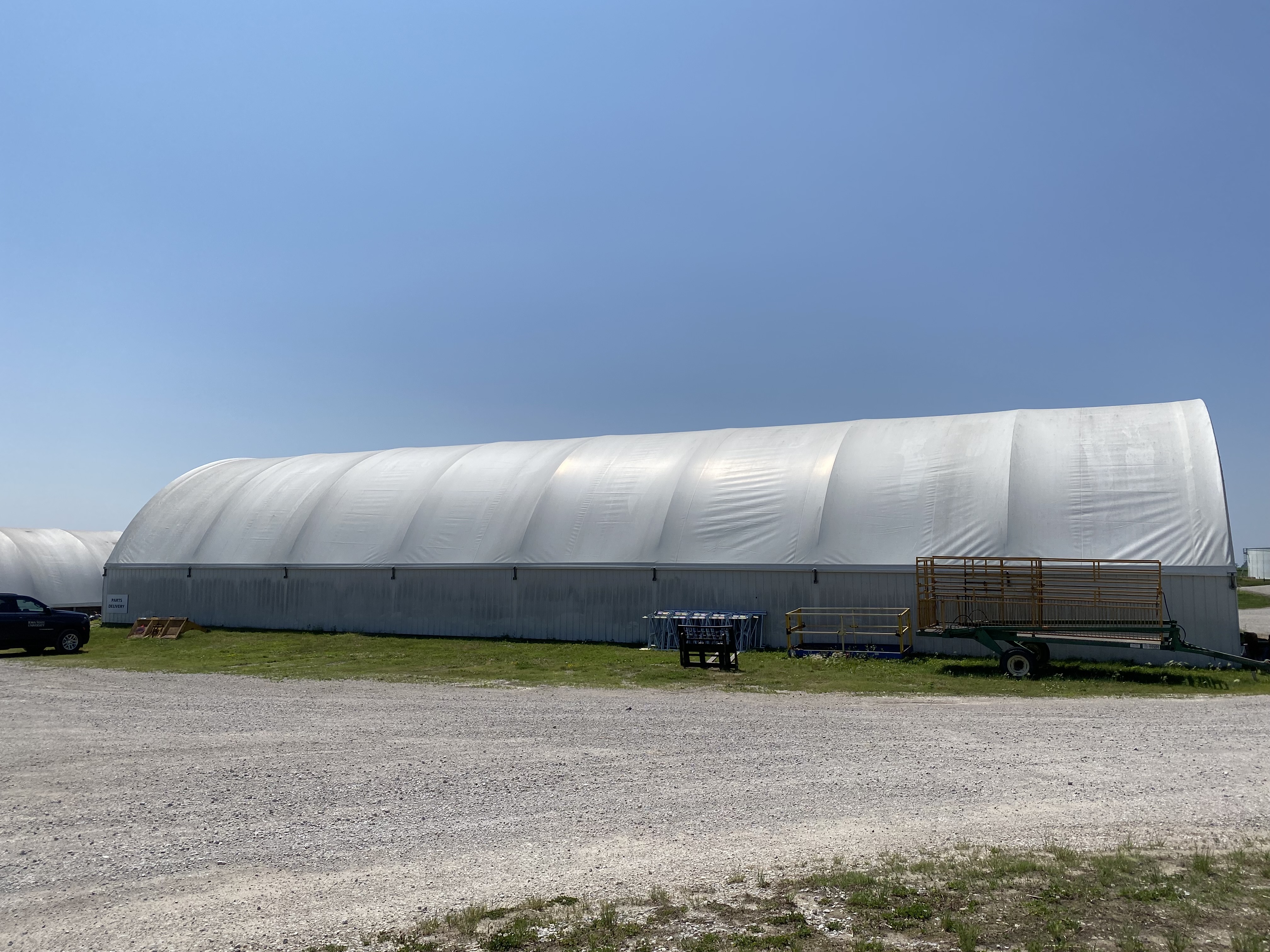}
        \caption{Hoop house.}
        \label{fig:hoop}
    \end{minipage}
    \hspace{0.1em}%
    \centering
    \begin{minipage}[htbp]{0.24\textwidth}
        \centering
        \includegraphics[scale=0.023]{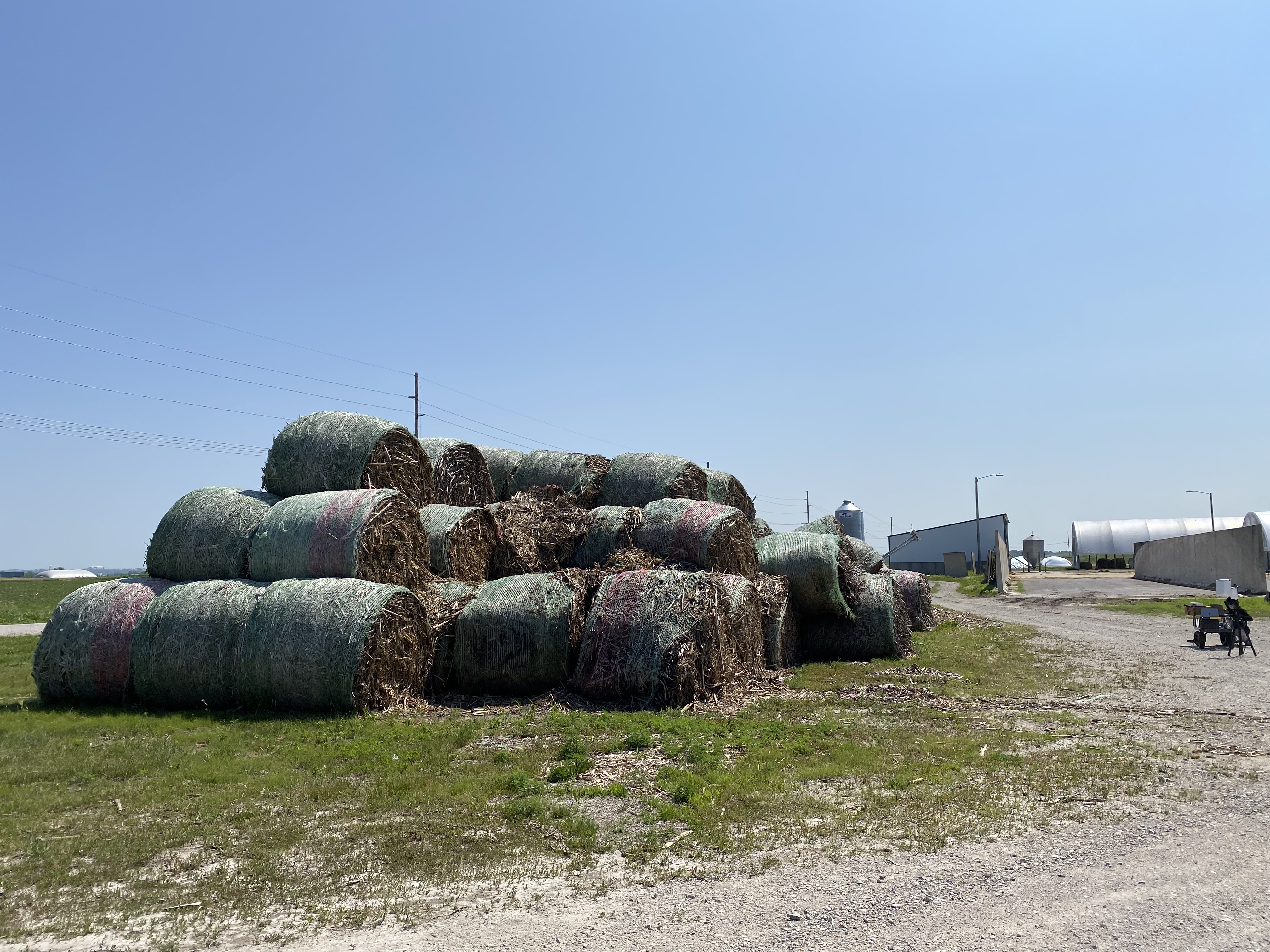}
        \caption{Hay pile.}
        \label{fig:Hay}
    \end{minipage}
    \hspace{0.1em}%
    \begin{minipage}[htbp]{0.24\textwidth}
        \centering
        \includegraphics[scale=0.023]{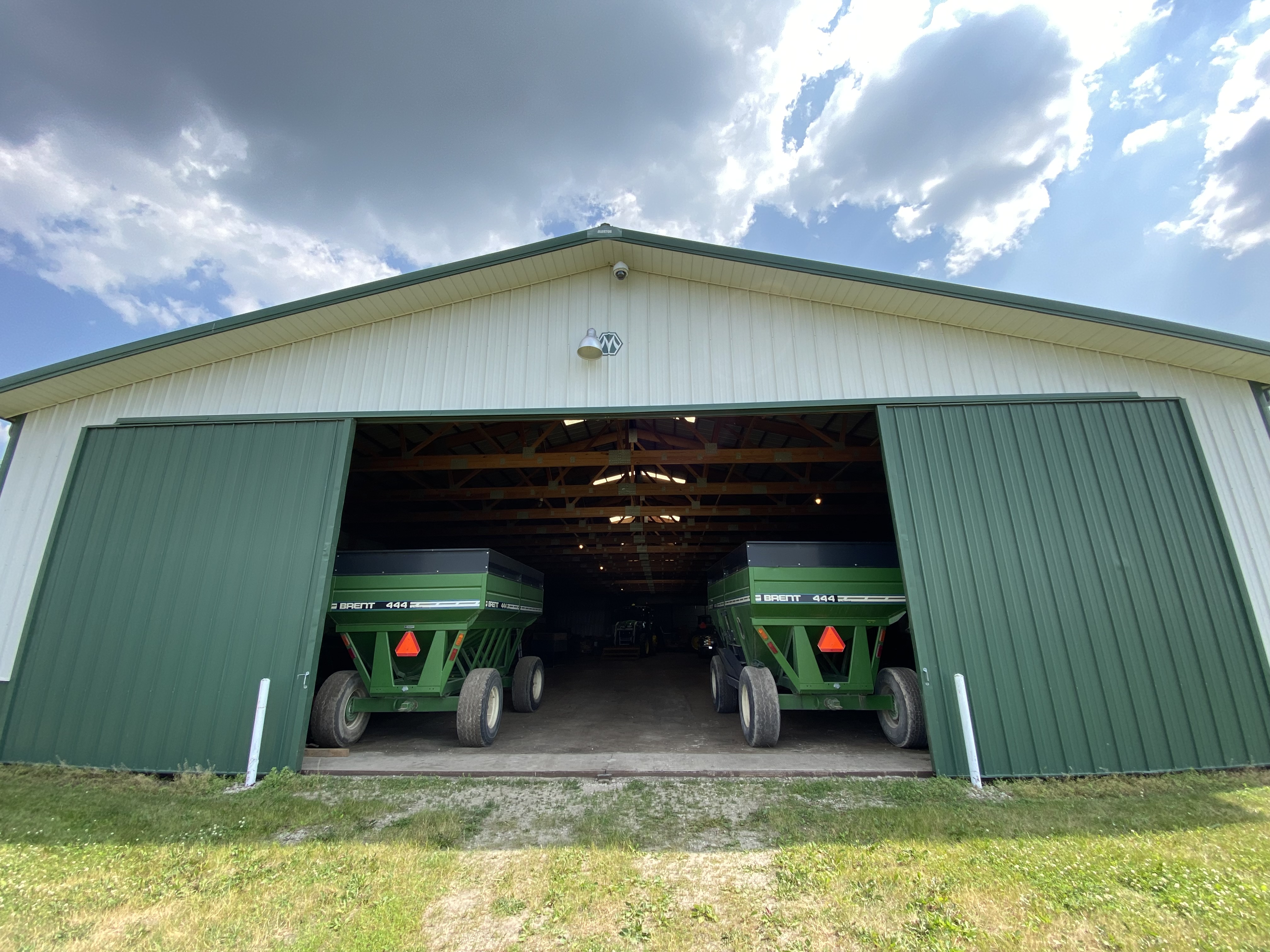}
        \caption{Storage building.}
        \label{fig:CurtissBuilding}
    \end{minipage}
    %\caption{Photos of agriculture buildings.}
    %\label{fig:buildings}
    \vspace{-0.3cm}
\end{figure*}

The blockage path loss results are listed in TABLE~\ref{tab:1}, while TABLE~\ref{tab:2} presents the results for the agricultural machinery storage building to demonstrate the impact of open and closed gates. To simplify the presentation of the results, we consider the measurements taken from the north side of the buildings as the baseline. The tables only show the additional path loss due to the buildings compared to the baseline. Moreover, Figs.~\ref{fig:barn} to~\ref{fig:CurtissBuilding} include snapshots of the buildings showing their shapes and structures.

\begin{table}[]
\footnotesize
    \centering
    \caption{Additional path loss due to obstruction by various farm buildings.}
    \resizebox{0.7\columnwidth}{!}{
    \begin{tabular}{| l | c | c |}
    \hline
        \textbf{Blockage Type} & \textbf{Mid-band} & \textbf{TVWS} \\
        \hline \hline
        Tree & 2~dB &  2~dB \\
        \hline
        Metal crop storage barn & 7~dB &  9~dB \\
        \hline
        Hoop house (Facing west) & 7~dB &  6~dB \\
        \hline
        Hoop house (Facing south) & 7~dB &  7~dB \\
        \hline
        Hay pile & 12~dB &  9~dB \\
         \hline
        Lactating barn (Outside-north) & 0~dB &  0~dB \\
         \hline
        Lactating barn (Inside) & 6~dB &  10~dB \\
         \hline
        Lactating barn (Outside-south) & 6~dB & $>$ 20~dB \\
        \hline
        Sheep barn (Outside-north) & 0~dB &  0~dB \\
         \hline
        Sheep barn (Inside-north) & 9~dB &  $>$ 10~dB \\
         \hline
        Sheep barn (Inside-middle) & 18~dB &  $>$ 10~dB \\
        \hline
        Sheep barn (Inside-south) & $>$ 18~dB & $>$ 10~dB \\
        \hline
        Sheep barn (Outside-south) & $>$ 18~dB & $>$ 10~dB \\
        \hline
    \end{tabular}}
    \label{tab:1}
    \vspace{-0.2cm}
\end{table}

From the findings in TABLE~\ref{tab:1}, it is observed that the hay piles cause significant signal blockage, which is somewhat unexpected. The orientation of the hoop house does not have a substantial impact. However, the orientation of the livestock barn does matter. For effective air circulation, the animal barns are open (and fenced) in the north-south direction or the east-west direction. The lactation barn, being north-south open, exhibits much less blockage compared to the sheep barn, which is east-west open.

\begin{table}[]
%\small
\footnotesize
    \centering
    \caption{Additional path loss due to obstruction by the agricultural machinery storage building.}
    \resizebox{0.8\columnwidth}{!}{
    \begin{tabular}{| l | c | c | c | c |}
    \hline
        \textbf{Location} & \textbf{North gate} &\textbf{South gate} &\textbf{Mid-band} & \textbf{TVWS} \\
        \hline \hline
        Outside-north & Close &  Close & 0~dB &  0~dB \\
        \hline
        Outside-north & Open &  Close & 0~dB &  0~dB \\
        \hline
        Inside-north & Close &  Close & 19~dB &  10~dB \\
        \hline
        Inside-north & Open &  Close & 9~dB &  5~dB \\
        \hline
        Inside-middle & Close &  Close & 18~dB &  10~dB \\
        \hline
        Inside-middle & Open &  Close & 9~dB &  6~dB \\
        \hline
        Inside-south & Close &  Close & 24~dB &  $>$ 10~dB \\
        \hline
        Inside-south & Open &  Close & 17~dB & $>$ 10~dB \\
        \hline
        Inside-south & Open &  Open & 16~dB &  10~dB \\
        \hline
        Outside-north & Close &  Close & $>$ 24~dB & $>$ 10~dB \\
        \hline
    \end{tabular}}
    \label{tab:2}
    \vspace{-0.4cm}
\end{table} 

TABLE~\ref{tab:2} presents additional path loss resulting from the obstruction caused by the agricultural machinery storage building. The table includes measurements conducted with different door configurations, as the building has two large gates facing north and south. Due to the presence of metal plates, the blockage effect is consistently significant, particularly when the north door is closed. However, the data indicates that the openness of the south door can slightly improve the signal quality due to the diffuse reflection of radio waves.
\vspace{-0.2cm}
\subsection{Dataset for Future Research}
ARA, as a large-scale multi-cell multi-band wireless experimental infrastructure, serves not only as a testbed for rural wireless and applications but also has the potential to play a unique role in providing valuable datasets to support various types of research. For example, one such research area could be AI-related research for channel modeling and channel occupancy prediction, where the channel information collected by the MIMO system and RF sensors, along with weather information collected by the weather station and disdrometer, can make significant contributions.

Currently, ARA is in the process of building a data warehouse to store the aforementioned data, and we plan to share it with the public through the ARA portal~\cite{loginOpenSt43:online} in the near future. The dataset generated from this measurement study will be included in the data warehouse.
%The measurement campaign dataset will primarily
The dataset consists of three parts: (1)~Skylark TVWS measurements, (2)~Ericsson Mid-band measurements, and (3)~weather information. All measurements are timestamped, enabling users to seamlessly integrate and analyze the data. Additionally, the data warehouse will also include measurement data that were collected during this measurement study but not yet analyzed and reported in this paper, such as data collected by the Skylark BS with multiple Skylark CPEs under different weather conditions. 

%there are supplementary data collected during the measurement campaign that were not presented in the paper, particularly the data collected by Skylark BS with multiple UEs under different weather conditions. Although these results are not included in this paper due to the introduction of numerous variables by multiple UEs, they still hold significant value.

%Furthermore, ARA is actively collecting wireless channel data under different extreme weather conditions, which can provide a rare and valuable wireless channel dataset. For example, we have already collected data during freezing rains in March 2023 and are currently investigating the impact of the wildfire smoke. We plan to continue gathering more data under various conditions to make further contributions to the academia and industry.
\vspace{-0.2cm}

\section{Conclusion}
\label{sec:Conclusion}
This work investigates the impact of weather conditions and agricultural buildings on TVWS and mid-band wireless channels in rural areas. Our study involved collecting wireless channel data during rainfall and analyzing the impact of rain rate and raindrop size. Our findings revealed that the rain rate has a more significant effect on signal attenuation than the raindrop size. Additionally, we discovered strong correlations between humidity, temperature, and path loss, suggesting a need for further exploration of the relationship between these three factors.
Another notable contribution of this paper is the inclusion of the path loss data resulting from agricultural buildings, which is an area of research that has only received limited attention so far. Furthermore, all data collected during this measurement study, including weather data, will be publicly accessible through the ARA portal~\cite{loginOpenSt43:online}. We anticipate that these datasets of real-world measurement results will prove valuable for estimation, modeling, and algorithm design pertaining to rural wireless channels.
%, by offering real-world measurement results.

%In future endeavors, our objective is to
%Our future plan includes gathering more data from rural areas to support diverse research avenues such as modeling and wireless channel prediction. We believe that this ongoing measurement study, in conjunction with our present findings, will serve as a crucial foundation for comprehending channel behavior in rural and agricultural settings, ultimately advancing the research on URLLC.
\vspace{-0.3cm}

\section*{Acknowledgment}

\vspace{-0.2cm}

This work is supported in part by the NIFA award 2021-67021-33775, and NSF awards 2130889, 2112606, 2212573, 2229654, 2232461.

\vspace{-0.4cm}

%\appendix

%\section*{Acknowledgment}
%This work is supported in part by the NIFA award 2021-67021-33775, and NSF awards 2130889, 2112606, 2212573, 2229654, and~2232461. We thank John Bennett-George and other ARA team members and partners for their contribution and support in the AraHaul deployment. 

\balance

\bibliographystyle{IEEEtran}
\bibliography{references}

%%%%%%%%%%%%%%%%%%%%%%%%%%%%%%%%%%%%%%%%%%%%%%%%%%%%%%%%%%%%%%%%%%%%%%%%%%%%%%%%
\end{document}